\documentclass[aps,prd,twocolumn,preprintnumbers,showpacs]{revtex4}

\usepackage{graphicx}
\usepackage{epsfig}
\usepackage{textcomp}
\usepackage{color}
\usepackage{url}

\def\theta{\vartheta}

\newcommand{\be}{\begin{equation}}
\newcommand{\ee}{\end{equation}}
\newcommand{\ba}{\begin{eqnarray}}
\newcommand{\ea}{\end{eqnarray}}
\newcommand{\lsim}   {\mathrel{\mathop{\kern 0pt \rlap
  {\raise.2ex\hbox{$<$}}}
  \lower.9ex\hbox{\kern-.190em $\sim$}}}
\newcommand{\gsim}   {\mathrel{\mathop{\kern 0pt \rlap
  {\raise.2ex\hbox{$>$}}}
  \lower.9ex\hbox{\kern-.190em $\sim$}}}

\begin{document}

\title{Are IceCube neutrinos unveiling PeV-scale decaying dark matter?}
\author{Arman Esmaili} 
\email{aesmaili@ifi.unicamp.br}
\affiliation{Instituto de Fisica Gleb Wataghin - UNICAMP, 13083-859, Campinas, SP, Brazil}
\author{Pasquale Dario Serpico} 
\email{serpico@lapth.cnrs.fr}
\affiliation{LAPTh, Univ. de Savoie, CNRS, B.P.110, Annecy-le-Vieux F-74941, France}

\begin{abstract}
Recent observations by IceCube, notably two PeV cascades accompanied by events at energies~$\sim (30-400)$~TeV, are clearly in excess over atmospheric background fluxes and beg for an astroparticle physics explanation. Although some models of astrophysical accelerators can account for the observations within current statistics, intriguing features in the energy and possibly angular distributions of the events make worth exploring alternatives. Here, we entertain the possibility of interpreting the data with a few PeV mass scale decaying  Dark Matter, with lifetime of the order of $10^{27}\,$s. We discuss generic signatures of this scenario, including its unique energy spectrum distortion with respect to the benchmark $E_\nu^{-2}$ expectation for astrophysical sources, as well as peculiar anisotropies. A direct comparison with the data show a good match with the above-mentioned features. We further discuss  possible future checks of this scenario.
\end{abstract}

\pacs{95.35.+d;   95.85.Ry	\hfill LAPTH-041/13}

\maketitle

\section{Introduction}\label{Intro}
A major recent discovery took place in neutrino astroparticle physics: the IceCube detector has reported the detection of two PeV neutrinos, a roughly $3\sigma$ excess above expected background rates~\cite{Aartsen:2013bka}, see also~\cite{Laha:2013lka} for an interpretative paper.  These events were found at the lower energy edge of an analysis optimized to search for cosmogenic neutrino cascade events, but the tentative signal does not have the rate and energy properties expected for such an origin. When the search criteria were extended to both cascade and track events of lower energies, an excess was found in the tens of TeV range~\cite{TeVexc}. By now, there is little doubt that this represents a discovery of a neutrino source of some kind, with evidence of the excess exceeding $4\sigma$. The most pressing and exciting issue remains to establish the nature of source(s).

In fact, the signal found has some intriguing features: while harder than the backgrounds, it does not seem to extend in energy much above a few PeV, otherwise  more energetic cascade events should have been detected. Also, albeit not yet statistically significant, there seems to be a deficit of events in the decade of energy just below the PeV pair, compared with expectations based on a simple power-law flux ranging from $\cal{O}$(10) TeV to PeV scale.

The arrival directions of the events also seem incompatible with an angular distribution peaked along the Galactic plane, while being compatible with an isotropic flux. Nonetheless, some enhancement toward inner Galaxy is not excluded. These features, together with some more general considerations, suggest that a significant fraction of events may originate in extragalactic sources, albeit presenting a  spectral softening or cutoff suppression at $E\lesssim 10^{17}\,$eV/nucleon primary energy. This can be obtained in realistic astrophysical models, like the starburst galaxies one in~\cite{Loeb:2006tw},  perhaps by stretching a bit the parameters. Some moderate Galactic component is anyway well consistent with the data and may be useful for diagnostics (a model of this  kind has been presented for example in~\cite{Fox:2013oza}).

All in all, astrophysical explanations for the observed flux are certainly viable. Yet, some intriguing features of the observed events leave room for alternative, non-standard physics interpretations. For example, in~\cite{Barger:2013pla}, the ``PeV excess'' was interpreted as s-channel enhancement of neutrino-quark scattering by a 0.6 TeV mass leptoquark. Also, in~\cite{Feldstein:2013kka}, the peculiar closeness in energy of the two PeV events was interpreted as due to a ``neutrino-line'' decay from heavy Dark Matter (DM) candidates.

Here, we explore the possibility that {\it all} of the above mentioned events (not only the couple of PeV cascades) may be due to DM decay. We point out that the case for a DM interpretation might be quite interesting for a number of reasons:

\begin{itemize}
\item[I]  The lack of signals of new physics at (sub-)TeV scale at LHC till now certainly weakens the case for ``WIMP'' candidates for DM. If one relaxes the ``theoretical prior'' for electroweak scale candidates, there is no reason to dismiss much heavier particles as an explanation of DM, lacking any other preferred energy scale. 
\item[II] The PeV range has a peculiar feature: it is natural to expect a {\it neutrino-first} detection, for reasons which have nothing to do with DM. Gamma rays have a Galactic-sized  mean free path at these energies, due to pair-production on the CMB. Electrons also lose energy very rapidly, thus leaving neutrinos as a channel of choice for indirect detection. In fact, for any candidate well above the TeV scale neutrinos offer significant advantages as probe~\cite{Esmaili:2012us,Murase:2012xs}. 
\item[III] As we shall detail below, the kind of spectral signal observed by IceCube may be {\it generic} for a wide class of unstable DM candidates with a mass in the right ballpark. Differently from what explored in~\cite{Feldstein:2013kka}, there is no need to single out very peculiar final states to explain the main features of the data. Also, we point out that DM decay predicts a very specific angular signal, which is qualitatively in agreement with what observed and may provide a further empirical test of the model, once further statistics will be collected.
\end{itemize}

The third item above requires to be made explicit. We devote to this purpose Sec.~\ref{signal} below. The rest of the article is composed as follows: In Sec.~\ref{results} we present our results for a few relevant cases; finally, in Sec.~\ref{conclusions} we discuss our findings and conclude.

\section{Dark Matter decay signal}\label{signal}
Here we closely follow~\cite{Esmaili:2012us} for methodology and notation. For more general notions on decaying DM models with an emphasis on their indirect detection prospects, see also the recent review~\cite{Ibarra:2013cra}, while for the earliest suggestion to use this kind of probe see~\cite{Gondolo:1991rn}. We just summarize the basic inputs below, while addressing to these references for more details. 

First, it is important to realize that the neutrino flux from DM decay has both a Galactic and an extragalactic contributions. Differently from the annihilating DM case, were the former term (Galactic) usually dominates on the latter (extragalactic), for decaying DM they are roughly comparable in flux, although different in angular distribution and in spectral shape.

The decay of DM particles in the Milky Way halo leads to the following differential flux:
\begin{equation}
  \frac{{\rm d} J_{\rm h}}{{\rm d}E_\nu}(l,b) = 
  \frac{1}{4\pi\,m_{\rm DM}\,\tau_{\rm DM}} 
  \frac{{\rm d}N_\nu}{{\rm d}E_\nu}
  \int_0^\infty {\rm d}s\; 
  \rho_{\rm h}[r(s,l,b)] \;,
  \label{halo-flux}
\end{equation}
where $m_{\rm DM}$ and $\tau_{\rm DM}$ are respectively the DM mass and lifetime; $\rho_{\rm h}(r)$ is the density profile of DM particles in our Galaxy as a function of distance from the Galactic center, $r$, and $dN_\nu/dE_\nu$ is the energy spectrum of neutrinos produced in the decay of a DM particle. The neutrino flux received at Earth depends on the Galactic coordinates, longitude $l$ and latitude $b$, and is given by a line-of-sight integral over the parameter $s$, which is related to $r$ by
\begin{equation}\label{galcoord}
  r(s,l,b) = \sqrt{s^2+R^2_\odot-2 s R_\odot \cos b\cos l}\,,
\end{equation}
where $R_\odot\simeq 8.5\,{\rm kpc}$ is the distance of the Sun to the Galactic center.

It is straightforward to calculate the average flux over the full sky, the result being
\begin{equation}\label{eq:halo}
\frac{{\rm d}J_{\rm h}}{{\rm d}E_\nu}=  D_{\rm h}  \frac{{\rm d}N_\nu}{{\rm d}E_\nu}\;.
\end{equation}
For our numerical analysis we will adopt a Navarro-Frenk-White density profile~\cite{Navarro:1996gj}
\begin{equation}
\rho_{\rm h}(r)\simeq \frac{\rho_h}{r/r_c (1+r/r_c)^2}\;,
\end{equation}
where $r$ is the distance to the Galactic center, $r_c\simeq20\,\text{kpc}$ is the critical radius and $\rho_h\simeq0.33\,{\rm GeV}\, {\rm cm}^{-3}$, which yields a DM density at the Solar System $\rho_\odot~=~0.39\,{\rm GeV}\, {\rm cm}^{-3}$~\cite{Catena:2009mf}. For this choice of parameters,
\begin{equation}\label{eq:Dhalo}
D_{\rm h}=1.7\times 10^{-12} 
\left(\frac{1\, {\rm PeV}}{m_{\rm DM}}\right)
\left(\frac{10^{27}\,{\rm s}}{\tau_{\rm DM}}\right)\,({\rm cm}^{2}\,{\rm s}\,{\rm sr})^{-1}\,.
\end{equation}
Note however that the decaying DM flux does not suffer from the large uncertainties related to the poor knowledge of the inner Galaxy profile. Lacking the quadratic enhancement in the signal proper of annihilating candidates, the role of the (better determined) local density is more prominent. We can thus anticipate little dependence of our conclusions on the specific choice made above.

In addition to the neutrino flux from decay of DM particles in the Milky Way halo, there is a second contribution stemming from the DM decays at cosmological distances, which produces an isotropic diffuse neutrino flux. The differential flux with respect to the received neutrino energy is given by:
\begin{equation}
\frac{{\rm d}J_{\rm eg}}{{\rm d}E_\nu} =
    \frac{\Omega_{\rm DM}\rho_{\rm c}}{4\pi m_{\rm DM} \tau_{\rm DM}}
    \int_0^\infty {\rm d}z\,
    \frac{1}{H(z)}
    \;\frac{{\rm d}N_\nu}{{\rm d}E_\nu}\left[(1+z)E_\nu\right],
  \label{eqn:NuFluxEG}
\end{equation}
where $H(z)=H_0 \sqrt{\Omega_\Lambda+\Omega_{\rm m}(1+z)^3}$ is the Hubble expansion rate as a function of redshift $z$ and $\rho_{\rm c}=5.5\times10^{-6}\,{\rm GeV}\, {\rm cm}^{-3}$ denotes the critical density of the Universe. Throughout this work we assume a $\Lambda$CDM cosmology with parameters $\Omega_\Lambda=0.6825$, $\Omega_{\rm m}=0.3175$, $\Omega_{\rm DM}=0.2685$ and $h\equiv H_0/100\,{\rm km}\,{\rm s}^{-1}\,{\rm Mpc}^{-1}=0.6711$, as derived from Planck temperature map data (see Table~2 in~\cite{Ade:2013zuv}). Similarly to Eq.~(\ref{eq:halo}), one can write
\begin{equation}
\frac{{\rm d}J_{\rm eg}}{{\rm d}E_\nu} =D_{\rm eg}  \int_0^\infty {\rm d}z\,
    \frac{{\rm d}N_\nu/{\rm d}E_\nu\left[(1+z)E_\nu\right]}{\sqrt{\Omega_\Lambda+\Omega_{\rm m}(1+z)^3}}\,,
  \label{eqn:NuFluxEG2}
\end{equation}
with
\begin{equation}\label{eq:Deg}
D_{\rm eg}=1.4\times 10^{-12} 
\left(\frac{1\, {\rm PeV}}{m_{\rm DM}}\right)
\left(\frac{10^{27}\,{\rm s}}{\tau_{\rm DM}}\right)\,({\rm cm}^{2}\,{\rm s}\,{\rm sr})^{-1}\,,
\end{equation}
which clearly suggest the similar magnitude of the Galactic and extragalactic signals.

Essentially, the three unknowns are $m_{\rm DM}$, $\tau_{\rm DM}$ and ${\rm d}N_\nu/{\rm d}E_\nu$. If, as in~\cite{Feldstein:2013kka}, one assumes that the DM decays basically in monochromatic neutrino line, the two PeV events fix both the mass and the lifetime, within factors of ${\cal O}$(1). However, for most DM models, a decay into neutrino lines is a sub-leading channel, with larger branching ratios are found in other SM final states. Hence, more generically it is the lower-energy continuum that fixes $\tau_{\rm DM}$, with the highest energy events fixing the branching ratio into hard (or monochromatic) neutrino channels (denoted $b_{\rm H}$ in the following), as well as the PeV-mass scale of the particle.

Remarkably, for a rough prediction of the spectral shape ${\rm d}N_\nu/{\rm d}E_\nu$ no detailed model of the DM sector is required. Phenomenologically, neutrino spectra from heavy particle decays/annihilations present both a hard and a soft component, denoted respectively with ``H'' and ``S''. The former one comes from final states containing ``primary'' neutrinos (such as a pair of neutrinos, $\nu\gamma$, etc.), but to some extent also other charged leptons, notably electron. On the other hand, other channels, and in particular those involving light quarks ($u\,, d\,,\,s$, here denoted as $q$), lead to significantly softer spectra. Hence we simply parameterize the spectrum as
\begin{equation}
\frac{{\rm d}N_\nu}{{\rm d}E_\nu} =(1-b_{\rm H})\left.\frac{{\rm d}N_\nu}{{\rm d}E_\nu}\right|_{\rm S}+b_{\rm H}\left.\frac{{\rm d}N_\nu}{{\rm d}E_\nu}\right|_{\rm H}\,.
  \label{eqn:modspectr}
\end{equation}
In the following section we will show some examples of spectra, specifying the ``S'' and ``H'' tree-level modes.

Additionally, it has been appreciated since more than a decade that  electroweak cascades are an important ingredient at center-of-mass energies of the order of PeV or larger~\cite{Berezinsky:2002hq}. Needless to say, also QCD parton jets play an important role, either because QCD final state is directly present (as in the soft channel of Eq.~(\ref{eqn:modspectr})) or, at very least, as byproduct of the electroweak cascades. In order to account for this effect, we rescale the results presented in~\cite{Cirelli:2010xx} for a 0.1 PeV annihilating DM---whose spectra are available numerically at~\cite{PPPC4DMID}---for few PeV candidates. Note that the ignorance of the DM model details, together with intrinsic theoretical error on the spectra prevents a precise calculation. Additionally, the current limited statistics would make it probably unnecessary. Hence, the spectra used should be  intended as ``educated guesses'', illustrative of the qualitatively expected shape, rather than detailed predictions.

Remarkably, as illustrated in the following section, with these generic ingredients and a minimal set of assumptions for DM physics, it appears rather easy to reproduce the ``unusual'' spectral shape hinted to by the data. This framework also lead to quite specific predictions, which can be falsified once sufficient statistics will be collected.

\section{Results}\label{results}

The galactic and extragalactic components of the neutrino flux at the Earth readily follow by plugging ${\rm d}N_\nu/{\rm d}E_\nu$ into Eq.~(\ref{eq:halo}) and Eq.~(\ref{eqn:NuFluxEG}), respectively. Our {\it benchmark case} is the choice ${\rm DM} \to \nu_e\bar{\nu}_e$ for the hard channel and ${\rm DM}\to q\bar{q}$ for the soft one. We also account for the neutrino mixing {\it en route} from production point to the Earth. Due to neutrino flavor oscillation, flux of $\nu_\alpha$ at Earth is $\Sigma_\beta P_{\alpha\beta} I_\beta$, where $P_{\alpha\beta}$ denotes probability of $\nu_\beta \to\nu_\alpha$ oscillation and $I_\beta$ represents the flux in $\beta$ flavor at the source. The observable flux are subject to complete decoherence, thus $P_{\alpha\beta}=\Sigma_i |U_{\alpha i}|^2 |U_{\beta i}|^2$, where $U_{\alpha i}$ represents the elements of PMNS mixing matrix, set here at the best-fit values from~\cite{GonzalezGarcia:2012sz}. Due to the oscillation of neutrinos, the flavor ratio of neutrino flux at the Earth from decaying DM is $J_e:J_\mu:J_\tau \simeq 1:1:1$, which is consistent with the observed numbers of muon-track and cascade events in IceCube.

\begin{figure}[t]
\centering 
\includegraphics[width=0.52\textwidth]{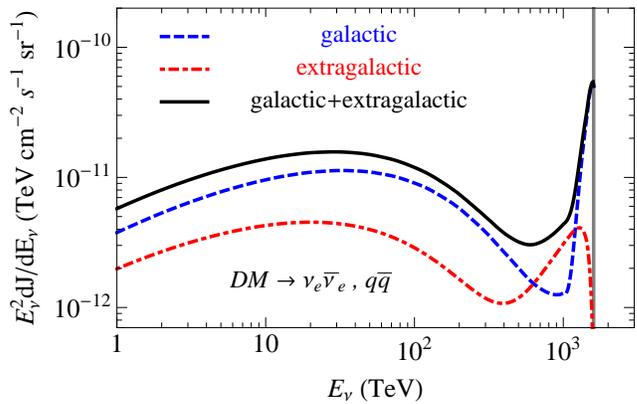}
\caption{\label{fig:spectrum}
The flux of neutrinos at the Earth form decaying DM with $m_{\rm DM}=3.2\,$PeV and $\tau_{\rm DM}=2\times 10^{27}\,$s and final states $\nu_e\bar{\nu}_e$ and $q\bar{q}$, with 12\% and 88\% branching ratios, respectively. The blue (dashed) and red (dot-dashed) curves are for galactic and extragalactic components, respectively. The black (solid) curves shows sum of the two components. The shown fluxes are $(\nu_e+\nu_\mu+\nu_\tau)/3$, including antineutrinos.}
\end{figure}

Fig.~\ref{fig:spectrum} shows the expected neutrino flux at Earth from decaying DM with $m_{\rm DM}=3.2$~PeV and $\tau_{\rm DM}=2\times 10^{27}\,$s, which as we will see gives a good fit to the IceCube data. The shown flux is the average of all neutrino and antineutrino flavors: $(\nu_e+\nu_\mu+\nu_\tau)/3$. The assumed DM mass stems from $m_{\rm DM}/2 \sim E_\nu^{\rm max}$, where $E_\nu^{\rm max}=1.6$~PeV is the maximum energy of observed events at IceCube; and $\tau_{\rm DM}$ is chosen in such a way to give two events in PeV range. The blue (dashed) and red (dot-dashed) curves correspond to galactic and extragalactic components, respectively; and the black solid curve for the sum of them. The gray vertical line shows the maximum energy of neutrino at $m_{\rm DM}/2$. For the branching ratio of hard channel DM decay (that is ${\rm DM}\to\nu_e\bar{\nu}_e$ for our benchmark), we assumed $b_{\rm H}=0.12$. The requested feature for the interpretation of IceCube data is clear from Fig.~\ref{fig:spectrum}: a peaked shape at $E_\nu\sim$~PeV accompanied by a dip in the range $\sim (0.3-1)$~PeV and populated spectrum below $\sim0.3$~PeV due to the softer $q\bar{q}$ channel (with cascade corrections) as well as the EW cascade tail from $\nu\bar{\nu}$. 

\begin{figure}[t]
\centering 
\includegraphics[width=0.48\textwidth]{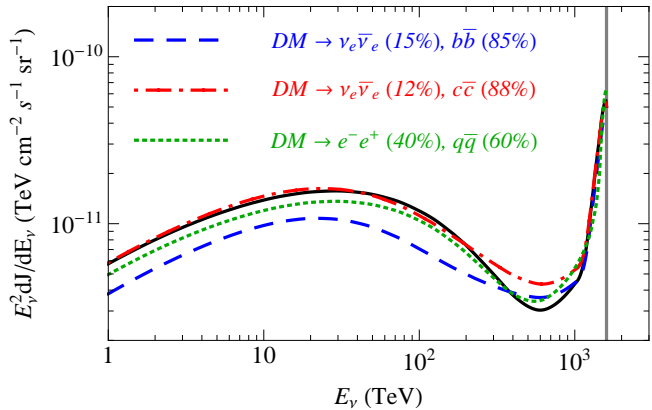}
\caption{\label{fig:Diffspectra}
The overall flux of neutrinos at the Earth for decaying DM to various channels. The black curve shows our benchmark ${\rm DM}\to\nu_e\bar{\nu}_e , q\bar{q}$ with 12\% and 88\% branching ratios, respectively. The blue (dashed), red (dot-dashed) and green (dotted) curves represent channels shown in legend with branching ratios in parentheses. The assumed values for $\tau_{\rm DM}$ are in the range $(1-3)\times 10^{27}\,$s. The shown flux is $(\nu_e+\nu_\mu+\nu_\tau)/3$, including antineutrinos.}
\end{figure}

The choice of final states sharing the qualitative features discussed above is by no means unique. In Fig.~\ref{fig:Diffspectra} we compare some alternative combinations of spectra presenting energy spectra similar to our benchmark decay channel (solid, black curve). In particular the soft channel in Eq.~(\ref{eqn:modspectr}) can be $b\bar{b}$ or $c\bar{c}$ final states and the hard channels can be replaced by $e^-e^+$ channel. As can be seen from Fig.~\ref{fig:Diffspectra}, the required shape of energy spectrum is recurring in all the shown channels. The $e^-e^+$ channel shows the importance of EW corrections (which are in fact quite large!): despite the fact that no hard neutrino channel is present at tree level, a sufficiently hard neutrino spectrum can be still obtained with a 40\% branching ratio in $e^-e^+$, thanks to the major role played by cascade radiation of massive gauge bosons (see~\cite{Kachelriess:2009zy,Ciafaloni:2010ti}). This fact may appear surprising, so we provide in the following a qualitative justification. First of all, even if one mostly radiates ``soft" gauge bosons, in a splitting process (say $e^- e^+ \to e^- W^+ \nu$) both the soft and the hard neutrino spectra are populated: the low-energy one via the soft (single or multiple) $W$ decay process and the high-energy one via the $\nu$'s which the electrons have converted into. Secondly, while naively these processes are suppressed by a power of $\alpha$ (weak fine structure) with respect to the three level, the presence of large logarithmic factor (of the type $\alpha \log (m_{\rm DM}^2/m_W^2)$) makes these ``corrections" sizable for massive particles, at the level of 10\% or larger of the tree-level result (for more technical details see e.g.~\cite{Ciafaloni:2010ti}). As a consequence, by varying both lifetime and branching ratio within a factor of only a few with respect to the naive fit obtained with the $\nu\bar{\nu}$ tree-level diagram, one is capable of fitting the spectrum even in the absence of tree-level neutrino emission. From the model building point of view, a DM decay to $e^-e^+$ and $\nu\bar{\nu}$ can be naturally constructed from the coupling of DM to the weak SU(2) lepton doublet $(\nu_\alpha,\ell_\alpha)$. For an equal decay branching ratio in the two components of the doublet, the corresponding modification of the parameters $\{\tau,b_H\}$ with respect to the pure $\nu\bar{\nu}$ case best fit parameters is thus less than a factor 2. Other choices for the final states (including for example massive gauge bosons, top quark and muon/tau leptons) would also produce spectra roughly compatible with observations, but for illustrative purposes in the following we shall concentrate on our benchmark case which presents the most marked differences with respect to a featureless power-law spectrum of astrophysical origin.

\begin{figure}[t]
\centering 
\includegraphics[width=0.535\textwidth]{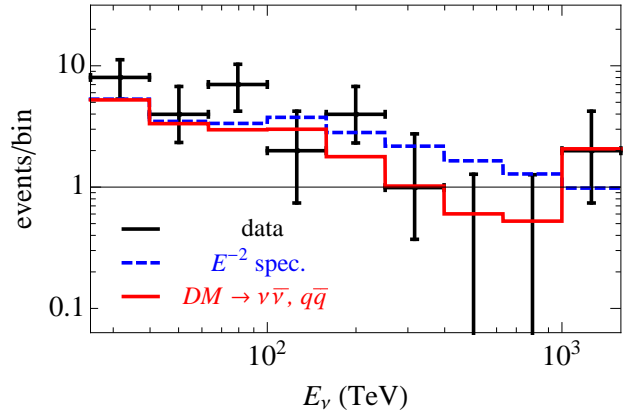}
\caption{\label{fig:events}
Comparison of the energy spectrum of observed events in IceCube with the expectations from DM decay with flux in Fig.~\ref{fig:spectrum} (red-solid) and generic $E_\nu^{-2}$ flux (blue-dashed). Both the observed events and predictions include background events due to atmospheric neutrinos and muons~\cite{TeVexc}. }
\end{figure}

The number of events at IceCube can be calculated by convoluting the flux at Earth with the exposure of the detector, such that the number of events in the bin $\Delta_i E_\nu$ is given by 
\begin{equation}\label{eq:events}
N_i = \int_{\Delta_i E_\nu} \left( \frac{{\rm d}J_{\rm h}}{{\rm d}E_\nu} + \frac{{\rm d}J_{\rm eg}}{{\rm d}E_\nu} \right) \mathcal{E}(E_\nu)~{\rm d}E_\nu~,
\end{equation} 
where for the exposure $\mathcal{E}$ we used the 662 days reported exposure in~\cite{Anchordoqui:2013qsi}. The result of our analysis is shown in Fig.~\ref{fig:events}. In this figure the red (solid) and blue (dashed) curves correspond to expected number of events from DM decay with the spectrum of Fig.~\ref{fig:spectrum} and a generic $E_\nu^{-2}$ spectrum, respectively; and the black points with error bars show the observed events. The following comments about Fig.~\ref{fig:events} are in order: 

\begin{itemize}
\item[1)] The branching ratio $b_{\rm H}=0.12$ of ${\rm DM}\to\nu_e\bar{\nu}_e$ is fixed mainly by requiring two PeV events, i.e. the last energy bin.

\item[2)] The DM lifetime $\tau_{\rm DM} = 2\times 10^{27}$~s is mainly determined by the low energy part of events. Let us mention that the assumed value of DM lifetime is compatible with the lower limit on $\tau_{\rm DM}$ obtained e.g. in~\cite{Esmaili:2012us} from the data of IceCube-22~\cite{Abbasi:2012cu}, but the two cannot be compared at face value. In fact, two issues should be taken into account: {\it i}) the lower limit in~\cite{Esmaili:2012us} is calculated with the assumption of $b_{\rm H}=1$, and as described there, the limit should be scaled for lower $b_{\rm H}$; {\it ii}) the monochromatic neutrino spectra used in~\cite{Esmaili:2012us} are sharper (and the bounds correspondingly stronger) than the ones used here. In particular, in~\cite{Esmaili:2012us} the EW corrections are not taken into account. EW corrections decrease the height of sharp line at $m_{\rm DM}/2$, as well as broadening it and introducing a smooth spectrum at low energy: as a consequence the lower limit weakens. Recalculating the lower limit on lifetime for the dataset of~\cite{Abbasi:2012cu} and the flux used in Eq.~(\ref{eq:events}) gives $\tau_{\rm DM}> 1.1\times 10^{26}\,$s (at 90\% C.L.) for $m_{\rm DM}=3.2\,$PeV, which is compatible with the assumed value in this paper. Our benchmark value is also consistent with the bounds derived in~\cite{Murase:2012xs}.

\item[3)] Since the maximum energy of each neutrino from DM decay is $m_{\rm DM}/2$, a sharp cut in the number of events exists for $E_\nu > m_{\rm DM}/2=1.6$~PeV, automatically matching the lack of observed events above the PeV {\it and} the two observed events at the PeV. For comparison, to accommodate a $E_\nu^{-2}$ spectrum with the excess from  ${\cal{O}}(10)\,$TeV to PeV scale, merely $\sim0.8$ events are expected in the high energy bin, while two have been measured~\footnote{Alternatively, an $E_\nu^{-2}$ flux producing 2 expected events in that bin would at the same time exacerbate the tension with the sub-PeV deficit and if extrapolated to high energies in absence of a cutoff, it would predict predict about seven events at supra-PeV energies (whereas none has been detected).} This upward fluctuation, while not significant (having a probability $\sim$~20\%) adds to the {\it three downward} fluctuations in the three preceding sub-PeV bins, where one event is observed whereas about 3 are expected, with a chance probability again of $\sim$~20\%. Since the highest energy observed events in IceCube consists of the two PeV events in the last bin of Fig.~\ref{fig:events}, a PeV scale DM decay interpretation naturally explains the data. Clearly, the DM mass and the resulting cut in energy can be chosen within the current uncertainty of the highest event energy. The main features of our results are robust with respect to the exact value of DM mass.

\item[4)] In the lower energy bins (below $\sim 200$~TeV), an $E_\nu^{-2}$ spectrum shows an agreement with the data comparable to (and actually a bit better than) our DM benchmark spectrum. But these bins are populated by both signal and background, with the latter (from atmospheric neutrinos and muons) being a more and more important component at lower energies (see~\cite{TeVexc}). For example, in the first few energy bins of Fig.~\ref{fig:events} almost half of the observed events can be interpreted as backgrounds. At $E_\nu\gtrsim200$~TeV, on the other hand, the contribution of background events should be negligible. In a sense, it is the last four more energetic bins that represent the most meaningful discriminator between an astrophysical and a DM model, with the latter showing a clearly better agreement. Of course, a typical astrophysical $E_\nu^{-2}$ spectrum is still in marginal agreement with the data due to the low statistics, but its goodness of fit (compared with alternative explanations) will be tested more meaningfully within few years by the increasing exposure of IceCube detector. 

\item[5)] The desired spectral feature from DM decay in Fig.~\ref{fig:spectrum} and the resulting energy distribution of events in Fig.~\ref{fig:events} can be roughly reproduced with a variety of choices for the particle physics parameters and is not tied to the specific value of $b_{\rm H}$ and decay channels used in Fig.~\ref{fig:events}, as suggested by Fig.~\ref{fig:Diffspectra}. Nonetheless, a DM mass in the range of $(2-10)\,$PeV and a lifetime to branching ratio factor $\tau_{\rm DM}/b_{\rm H}\sim  10^{28}\,$s are rather generic, at least to reproduce the highest energy data.
\end{itemize}

A unique and predictive feature of DM decay interpretation of IceCube data, which is {\it independent} of the exact shape of the spectrum, is the expected anisotropy in observed events. Due to the off-center position of the Solar System in Milky Way, the halo component of the flux in Eq.~(\ref{halo-flux}) is larger in the direction of Galactic center (that is $l=b=0$), see for example Fig. A1 in~\cite{Serpico:2009vz}. Interestingly, the IceCube data also show an excess in this direction, which corresponds to $\delta\approx-29^\circ$, where $\delta$ is the declination (see~\cite{TeVexc}). A detailed calculation of expected anisotropy in the case of DM decay requires information about direction dependence of exposure, which is not provided by the IceCube collaboration. For illustrative purposes, however,  note that about 20\% of the Galactic signal (or about 15\% of the total one) should come from the inner 30$^\circ$ around the Galactic center, as opposed to less than 7\% in the case of an isotropic flux. Although these differences are probably too small to be probed with current statistics, significant diagnostic potential should be available in a decade or so of collecting time. Note that this test would work even in the absence of a hard spectrum near the endpoint, i.e. $b_H\to 0$. Even a featureless DM spectrum dominated by soft neutrinos, in fact, would have the same angular characteristics described above.

\section{Discussion and Conclusion}\label{conclusions}
Over the last year, evidence has been building up about the existence of a TeV-to-PeV diffuse neutrino flux, on the top of known backgrounds.  However, the interpretation of the current signal is not yet clear. 

Here, we argued that some features are sufficiently intriguing that an explanation in terms of decaying DM seems worth exploring in more details. We showed that assuming that such a particle exists, with a lifetime  allowed by present constraints and with reasonable final state spectra, it could fit quite easily the unusual energy-shape of the excess: in general, the sum of a hard (e.g. leptonic) and a soft (e.g. hadronic) final state, even accounting for EW as well as QCD cascade corrections, provides a much closer match to the measured shape of the excess than a simple astrophysical power-law close to $E_\nu^{-2}$. A better determination of the spectrum of the excess, in particular the behavior above $E_\nu\sim (1-2)\,$PeV and in the decade just below it, $E_\nu\gtrsim (0.1-0.2)\,$PeV, should provide further clues in this sense: a DM explanation {\it requires} that a rapid drop~\footnote{The flux might not drop to zero, of course, if the DM signal is accompanied by some subleading astrophysical background. In this case the cutoff would turn into a spectral discontinuity.} of the signal is observed above the endpoint of the spectrum, which must be below (or equal to) half of the DM particle mass.  Additionally, a sub-PeV dip in the spectrum is frequently obtained, albeit its prominence  depends somewhat on the exact final state channels.

Also, we commented on the fact that this DM explanation has very peculiar predictions about the {\it angular} distribution of the events, which may be discriminated against either an isotropic flux or a Galactic-plane enhanced flux, once sufficient statistics will be collected. For this task, the more numerous lower-energy events in the tens of TeV range---which are less effective for spectral shape diagnostics---may prove essential.

Finally, we note that the neutrino flux is generically accompanied by a comparable flux of photons, whose extragalactic component is however degraded below $E_\gamma \sim {\cal O}(100)\,$GeV via electromagnetic cascades (pair-production and inverse Compton events) initiated onto the extragalactic background light and the CMB. Since the DM number density scales as $1/m_{\rm DM}$, but the injected energy goes as $m_{\rm DM}$, as long as most of the photons emitted initiate the cascades, the bound is basically calorimetric and these constraints on $\tau_{\rm DM}$ are almost mass-independent. Existing studies quantify them in the range of $10^{26}-10^{27}\,$s (see for example~\cite{Murase:2012xs} and~\cite{Cirelli:2012ut}), so certainly compatible with lifetime values required here, but not wildly larger. It is likely that a non-negligible fraction (say, larger than ${\cal O}(10\%)$) of the diffuse flux at tens or hundreds of GeV may be due to these DM secondary photons. In the future, a better understanding of this flux, both from the experimental and the modeling side, may thus open a further possibility of diagnostics.

In conclusion, opening the TeV-PeV astronomy window would represent, by itself, a seminal achievement for astrophysics. However, this will likely have implications for fundamental physics as well: For example, it has already been argued that proving the existence of some extragalactic component at PeV energies would also imply a big improvement on bounds on Lorentz violation in the neutrino sector~\cite{Borriello:2013ala}. Eventually, it might well be that the recent discovery by IceCube will offer a serendipitous solution to the long quest for DM identification.


\begin{acknowledgments}
At LAPTh, this activity was supported by the Labex grant ENIGMASS and the ANR DMAstroLHC (ANR-12-BS05-0006- 01). A.~E. thanks LAPTh for hospitality, where this work was partially done. A.~E. acknowledges financial support by the funding grant 2009/17924-5, from S\~ao Paulo Research Foundation (FAPESP). The authors thank Alexei~Yu.~Smirnov for his comments on the manuscript.

\end{acknowledgments}



\begin{thebibliography}{99}

\bibitem{Aartsen:2013bka} 
  M.~G.~Aartsen {\it et al.}  [IceCube Collaboration],
  arXiv:1304.5356 [astro-ph.HE].
  
\bibitem{Laha:2013lka} 
  R.~Laha, J.~F.~Beacom, B.~Dasgupta, S.~Horiuchi and K.~Murase,
  arXiv:1306.2309 [astro-ph.HE].
  
\bibitem{TeVexc}

https://events.icecube.wisc.edu/getFile.py/access?\\contribId=76\&sessionId=41\&resId=0\&materialId=\\slides\&confId=46

\bibitem{Loeb:2006tw} 
  A.~Loeb and E.~Waxman,
  JCAP {\bf 0605}, 003 (2006)
  [astro-ph/0601695].

\bibitem{Fox:2013oza} 
  D.~B.~Fox, K.~Kashiyama and P.~Meszaros,
  arXiv:1305.6606 [astro-ph.HE].
    
\bibitem{Barger:2013pla} 
  V.~Barger and W.~-Y.~Keung,
  arXiv:1305.6907 [hep-ph].

\bibitem{Feldstein:2013kka} 
  B.~Feldstein, A.~Kusenko, S.~Matsumoto and T.~T.~Yanagida,
  arXiv:1303.7320 [hep-ph].
  
  
\bibitem{Esmaili:2012us} 
  A.~Esmaili, A.~Ibarra and O.~L.~G.~Peres,
  JCAP {\bf 1211}, 034 (2012)
  [arXiv:1205.5281 [hep-ph]].

  
\bibitem{Murase:2012xs} 
  K.~Murase and J.~F.~Beacom,
  JCAP {\bf 1210}, 043 (2012)
  [arXiv:1206.2595 [hep-ph]].
  
  
  \bibitem{Ibarra:2013cra} 
  A.~Ibarra, D.~Tran and C.~Weniger,
  arXiv:1307.6434 [hep-ph].
  
\bibitem{Gondolo:1991rn} 
  P.~Gondolo, G.~Gelmini and S.~Sarkar,
  Nucl.\ Phys.\ B {\bf 392}, 111 (1993)
  [hep-ph/9209236].

\bibitem{Navarro:1996gj}
  J.~F.~Navarro, C.~S.~Frenk and S.~D.~M.~White,
  Astrophys.\ J.\  {\bf 490} (1997) 493
  [astro-ph/9611107].

\bibitem{Catena:2009mf}
  R.~Catena and P.~Ullio,
  JCAP {\bf 1008} (2010) 004
  [arXiv:0907.0018 [astro-ph.CO]].
  
  
\bibitem{Ade:2013zuv} 
  P.~A.~R.~Ade {\it et al.}  [Planck Collaboration],
  arXiv:1303.5076 [astro-ph.CO].
    
\bibitem{Berezinsky:2002hq} 
  V.~Berezinsky, M.~Kachelriess and S.~Ostapchenko,
  Phys.\ Rev.\ Lett.\  {\bf 89}, 171802 (2002)
  [hep-ph/0205218].

  
\bibitem{Cirelli:2010xx} 
  M.~Cirelli, G.~Corcella, A.~Hektor, G.~Hutsi, M.~Kadastik, P.~Panci, M.~Raidal and F.~Sala {\it et al.},
  JCAP {\bf 1103}, 051 (2011)
  [Erratum-ibid.\  {\bf 1210}, E01 (2012)]
  [arXiv:1012.4515 [hep-ph]].
  
  \bibitem{PPPC4DMID}
\texttt{http://www.marcocirelli.net/PPPC4DMID.html}

\bibitem{GonzalezGarcia:2012sz} 
  M.~C.~Gonzalez-Garcia, M.~Maltoni, J.~Salvado and T.~Schwetz,
  JHEP {\bf 1212}, 123 (2012)
  [arXiv:1209.3023 [hep-ph]].


\bibitem{Borriello:2013ala} 
  E.~Borriello, S.~Chakraborty, A.~Mirizzi and P.~D.~Serpico,
  Phys.\  Rev.\  D 87, {\bf 116009} (2013)
  [arXiv:1303.5843 [astro-ph.HE]].
  
\bibitem{Anchordoqui:2013qsi} 
  L.~A.~Anchordoqui, H.~Goldberg, M.~H.~Lynch, A.~V.~Olinto, T.~C.~Paul and T.~J.~Weiler,
  arXiv:1306.5021 [astro-ph.HE].
  
\bibitem{Abbasi:2012cu} 
  R.~Abbasi {\it et al.}  [IceCube Collaboration],
  Phys.\ Rev.\ D {\bf 86}, 022005 (2012)
  [arXiv:1202.4564 [astro-ph.HE]].

 \bibitem{Kachelriess:2009zy} 
  M.~Kachelriess, P.~D.~Serpico and M.~A.~.Solberg,
  Phys.\ Rev.\ D {\bf 80}, 123533 (2009)
  [arXiv:0911.0001 [hep-ph]].
  
 \bibitem{Ciafaloni:2010ti} 
P.~Ciafaloni, D.~Comelli, A.~Riotto, F.~Sala, A.~Strumia and A.~Urbano,
  JCAP {\bf 1103}, 019 (2011)
  [arXiv:1009.0224 [hep-ph]].
  
\bibitem{Serpico:2009vz} 
  P.~D.~Serpico and D.~Hooper,
  New J.\ Phys.\  {\bf 11}, 105010 (2009)
  [arXiv:0902.2539 [hep-ph]].
  
\bibitem{Cirelli:2012ut} 
  M.~Cirelli, E.~Moulin, P.~Panci, P.~D.~Serpico and A.~Viana,
  Phys.\ Rev.\ D {\bf 86}, 083506 (2012)
  [arXiv:1205.5283 [astro-ph.CO]].
  
  
\end{thebibliography}
\end{document}